\documentclass[aps,twocolumn,superscriptaddress]{revtex4-1}
\usepackage{amsmath}
\usepackage{amssymb}
\usepackage{graphicx}
\usepackage{epstopdf}
\usepackage[colorlinks=true]{hyperref}

\newcommand{\tr}{\mathrm{Tr}}
\usepackage{physics}
\usepackage{mathrsfs}
\usepackage{comment}

\renewcommand\({\begin{equation}}	
\renewcommand\){\end{equation}}
\begin{document}

\title{Tuning entanglement by squeezing magnons in anisotropic magnets}

\author{Ji Zou}
\affiliation{Department of Physics and Astronomy, University of California, Los Angeles, California 90095, USA}
\author{Se Kwon Kim}
\affiliation{Department of Physics and Astronomy, University of California, Los Angeles, California 90095, USA}
\affiliation{Department of Physics and Astronomy, University of Missouri, Columbia, Missouri 65211, USA}
\author{Yaroslav Tserkovnyak}
\affiliation{Department of Physics and Astronomy, University of California, Los Angeles, California 90095, USA}

\begin{abstract}
We theoretically study the entanglement between two arbitrary spins in a magnetic material,  where magnons naturally form a general squeezed coherent state, in the presence of an applied magnetic field and axial anisotropies. Employing concurrence as a measure of entanglement, we demonstrate that spins are generally entangled  in thermodynamic equilibrium, with the amount of  entanglement controlled by the external fields and anisotropies. As a result, the magnetic medium can serve as a resource to store  and process quantum information. We, furthermore, show that the entanglement can jump discontinuously when decreasing the transverse magnetic field. This tunable entanglement can  be potentially used as an efficient switch in quantum-information processing tasks.
\end{abstract}

\date{\today}
\maketitle

\section{Introduction}

Entanglement \cite{RevModPhys.80.517, Plbnio:2007:IEM:2011706.2011707} is a measure of how much quantum information is stored in a quantum state and is one of  the most fundamental properties that distinguish a quantum phenomenon from its classical counterpart. It was under a severe skepticism, however, since the discovery of quantum mechanics  \cite{PhysRev.47.777}, due to its nonlocality that appeared to violate the local-realism view of causality. Following the derivation of Bell's inequalities, which rendered the nonlocal features of quantum theory accessible to experimental verification, numerous experiments in different systems have been carried out, including photons, neutrinos, electrons, molecules as large as buckyballs, and even small diamonds, unequivocally demonstrating the existence of quantum entanglement \cite{PhysRevLett.75.4337, Zhao:2004aa, Yao:2012aa, PhysRevLett.117.050402, Hensen:2015aa, Arndt:1999aa, Lee1253}.  In parallel with these developments, quantum entanglement has come to be recognized as a valuable resource in quantum-information processing \cite{nielsen}. For example, a quantum computer can be much faster and more powerful than a classical one for certain computational tasks, by taking the advantage of the superposition and entanglement in a quantum system \cite{Feynman1982}. Moreover, we can realize several quantum protocols, such as  teleportation, exclusively with the help of entangled states \cite{PhysRevLett.70.1895}.  These merits of entanglement in quantum information science stimulate the research trying to coherently  prepare and manipulate it in various systems.

A magnon Bose-Einstein condensate (BEC) \cite{Bunkov:2018aa}, where quanta of spin waves condense into a single state, may be taken as a platform to look for  controllable entanglement \cite{PhysRevA.66.052323}, since particles in a condensate are distributed over space in a coherent way. Magnon BEC is attractive in practice, as it can be driven by microwave \cite{Demokritov:2006aa, PhysRevLett.99.037205, PhysRevLett.100.047205, PhysRevLett.101.257201, PhysRevB.80.060401, PhysRevB.79.174411, Rezende:2010aa} or electronic \cite{PhysRevLett.108.246601, PhysRevB.90.094409} pumping in an insulating ferromagnet through a quasiequilibration process at room temperature. Without magnon pumping, we can also achieve magnon BEC in equilibrium, by introducing an easy-plane anisotropy in the magnetic system \cite{PhysRevLett.116.117201}. As we show in this paper, magnons condense by forming a general squeezed coherent state, when the system is subjected to external magnetic fields and anisotropies. A squeezed coherent state \cite{nicequanoptics}, akin to a coherent state, is  a minimum uncertainty state but with uncertainties of conjugate operators being different. This state has been  investigated extensively in quantum optics, resulting in many applications. For example, it can be used to improve the precision of atom clocks \cite{PhysRevLett.104.250801,Louchet_Chauvet_2010} and quantum-information processing in the continuous-variables regime \cite{RevModPhys.77.513}. 

In this paper, we study the entanglement of arbitrary two spins in a magnetic system (assuming the number of spin sites is $N_0$), utilizing the concurrence $\mathcal{C}$ \cite{PhysRevLett.80.2245,Plbnio:2007:IEM:2011706.2011707} as a measure of entanglement, where magnons are condensed into a general squeezed coherent state. The average number of condensed magnons in such a state can be tuned by the external field and magnetic anisotropies \cite{nicequanoptics}. We distinguish between two types of magnons: coherent magnons related to a uniform order-parameter tilting and squeezed magnons related to the anisotropic squeezing effect [see Eq.~(\ref{number}) below]. From numerical  analysis, we find that the system transits abruptly to a highly entangled state from an unentangled state, when we decrease the number of coherent magnons (denoted by $N_c$) across a critical value that is determined by the number of squeezed magnons (denoted by $N_s$):
\(N_c=\sqrt{2N_0N_s}.  \)
This can be potentially used as a switch in quantum-information processing tasks. Whereas a simple coherent state has no entanglement, a squeezed vacuum state is entangled with concurrence
 \( \mathcal{C}=\frac{2}{N_0}\frac{\sqrt{N_s}}{\sqrt{N_s+1}+\sqrt{N_s}}. \)
This concurrence will increase as the number of squeezed magnons rises.  Resembling squeezed light in quantum optics \cite{RevModPhys.77.513,Furusawa706,PhysRevLett.101.130501}, our squeezed coherent  magnetic  states can also serve as an essential resource to realize continuous-variables protocols for quantum communication, unconditional quantum teleportation, and one-way quantum computing. Apart from this, we also discuss the entanglement when the condensate is in a Fock state and we match our result with the entanglement of Dicke states \cite{Friedberg_2007}, which has been well understood in quantum spin squeezing \cite{MA201189,PhysRevA.68.033821,Wang2002}. In contrast to the entanglement between macroscopic building blocks of cavity magnetomechanical systems \cite{Li_2019,PhysRevResearch.1.023021}, we are considering the intrinsic spin entanglement within a quantum medium.

The paper is structured as follows: In Sec.~\ref{model}, we introduce the model and discuss the ground state in terms of coherent squeezing. In Sec.~\ref{measure}, an entanglement measure, known as the entanglement formation is briefly reviewed, along with its relation to concurrence \cite{PhysRevLett.80.2245}. In Sec.~\ref{main}, we derive main results of this paper, namely, the entanglement of a general squeezed coherent magnetic state, including its two special cases: coherent states and squeezed vacuum states. The entanglement of Fock states and the distance dependence of  entanglement in the thermodynamic limit are also examined. A summary and outlook are offered in Sec.~\ref{summary}.

\section{Model}
\label{model}
Our model system is a set of localized spins interacting through a nearest-neighbor exchange coupling on a three-dimensional lattice, with axial anisotropies and a tilted magnetic field, according to the following Hamiltonian \cite{hamiltonian}:
\begin{eqnarray}
H&=&-J\sum_{\langle i,j \rangle }\vb{S}_i\cdot \vb{S}_{j} -\vb{B}\cdot \sum_i\vb{S}_i+H_1+H_2, \label{totalH}\\
H_1&=&\frac{K}{w}\sum_{\langle i,j \rangle}(S_i^xS_{j}^x-S_i^yS_{j}^y   ), \label{H1}\\
H_2&=& -\vb{h}\cdot \sum_i\vb{S}_i.
\end{eqnarray}
Here, $\vb{S}_i=\boldsymbol{\sigma}_i/2$ is the spin operator on the $i$th site ($\boldsymbol{\sigma}$ are Pauli matrices and  we have set $\hbar=1$ for simplicity), $w$ is the lattice coordination number (for example, $w=6$ for a simple cubic lattice), $J>0$ is the exchange constant of a simple Heisenberg magnet ($J\gg B,|K|$), $|K|$ is the anisotropy strength, $\langle i,j \rangle$ stands for all nearest-neighbor pairs, and $\vb{B}=B\,\hat{\vb{z}}$, $\vb{h}=h\,\hat{\vb{x}}$ are the external fields (absorbing all constant factors). We will restrict our discussion to the case $B>|K|$, so that quantum spin fluctuations can be expanded around the $z$ axis.

We will focus on the low-temperature limit, $T\ll J$, where thermal magnons are dilute.  It is convenient to switch from the $\text{SU}(2)$ spin algebra to the bosonic algebra: $S^+_i= a_i, S^z_i=1/2-a^\dagger_ia_i$, where $a^\dagger_i, a_i$ are the magnon creation and annihilation operators in real space that obey bosonic commutation relations. This transformation is exact, when complemented with the hard-core repulsion for magnons \cite{hardcore}. In the dilute limit, where the average magnon density is small $\langle a_i^\dagger a_i\rangle \ll 1$, as in our case of $J\gg B>|K|$, we can relax the hard-core repulsion constraint and, as a result, the Hamiltonian $H$ can be linearized and rewritten as

\begin{multline}
H=\sum_{\vb{k}}(2J \vb{k}^2+B) a^\dagger_{\vb{k}} a_{\vb{k}}+ \frac{K}{2} \sum_{\vb{k}}(a^\dagger_{-\vb{k}} a^\dagger_{\vb{k}}+a_{\vb{k}}a_{-\vb{k}})\\
-\frac{h\sqrt{N_0}}{2}(a^\dagger +a)+\cdots.
\end{multline}
Here, $a\equiv a_{q=0}$, the ellipsis represents nonlinear terms, and $N_0$ is the total number of sites in the system. Note we have set the lattice constant to be $1$, which means all quantities with length dimension will be measured in unit of the lattice constant.   The above Hamiltonian can be diagonalized by applying  Bogoliubov transformations \cite{Bogoljubov:1958aa,Kamra:2016aa} and rewritten as 
\(H=\sum_{\vb{k}\neq 0} \omega(\vb{k}) b^\dagger_{\vb{k}} b_{\vb{k}} + \omega b^\dagger b, \)
where $b^\dagger, b$ and $b^\dagger_{\vb{k}}, b_{\vb{k}}$ are  bosonic operators, $\omega(\vb{k})=\sqrt{(2J \vb{k}^2+B)^2-K^2 }$ and $\omega\equiv \omega(\vb{k}=0)$. The operator $b$ and $b_{\vb{k}}$ are related to $a$ and $a_{\vb{k}}$  via 
\begin{eqnarray}
b&=&D(\alpha)S(r)a S^\dagger(r)D^\dagger (\alpha);\\
b_{\vb{k}}&=&S(\phi_{\vb{k}}/2)a_{\vb{k}}S^\dagger(\phi_{\vb{k}}/2), \, \vb{k}\neq 0.
\end{eqnarray}
$S(\phi_{\vb{k}}/2)=e^{(a_{\vb{k}}a_{-\vb{k}}-a^\dagger_{\vb{k}}a^\dagger_{-\vb{k}})\phi_{\vb{k}} /2}$ is a two-mode squeezing operator \cite{Yun_Xia_2008,PhysRevA.63.022305} and $\phi_{\vb{k}}$ is determined by $\tanh\phi_{\vb{k}}=K/(2J \vb{k}^2+B)$. $S(r)=e^{[ a^2- (a^\dagger)^2]r/2}$ is a squeezing operator \cite{nicequanoptics} with $r=\phi_{\vb{k}=0}/2$. $D(\alpha)=e^{\alpha a^\dagger-\alpha^* a}$ is a displacement operator with $\alpha=h\sqrt{N_0}e^{-2r}/2\omega$.

The ground state $\ket{\psi}$ is given by $b\ket{\psi}=0, b_{\vb{k}}\ket{\psi}=0$ for all $\vb{k}$ and thus
\( \ket{\psi}=\big(D(\alpha)S(r)\ket{0}\big)\otimes \bigg( \prod_{\vb{k}\neq 0} S(\phi_{\vb{k}}/2)\ket{0} \bigg), \label{state}  \)
where $\ket{0}$ is the Fock vacuum defined by $a_{\vb{k}}\ket{0}=0$ for any wavenumber $\vb{k}$. The average number of magnons in the ground state is 
\(  \langle a^\dagger_{\vb{k}} a_{\vb{k}} \rangle =\delta_{\vb{k},0}|\alpha|^2+\sinh^2\frac{\phi_{\vb{k}}}{2}. \label{number} \)
In the large exchange-coupling limit with the size of the system being finite, the effect of nonzero wavenumber modes is negligible and the ground state reduces to the so called squeezed coherent state $\ket{\psi}=D(\alpha)S(r)\ket{0}$. Under the circumstances, $ \langle a^\dagger_{\vb{k}} a_{\vb{k}} \rangle =\delta_{\vb{k},0}(|\alpha|^2+\sinh^2r)$. We refer to the part related to the coherent parameter $\alpha$  as coherent magnons and the part related to  the squeezing parameter $r$ as squeezed magnons, denoted by 
\(N_c\equiv |\alpha|^2\,\, \text{and}\,\, N_s\equiv \sinh^2r,\) 
respectively.

\begin{figure}
\includegraphics[scale=0.23]{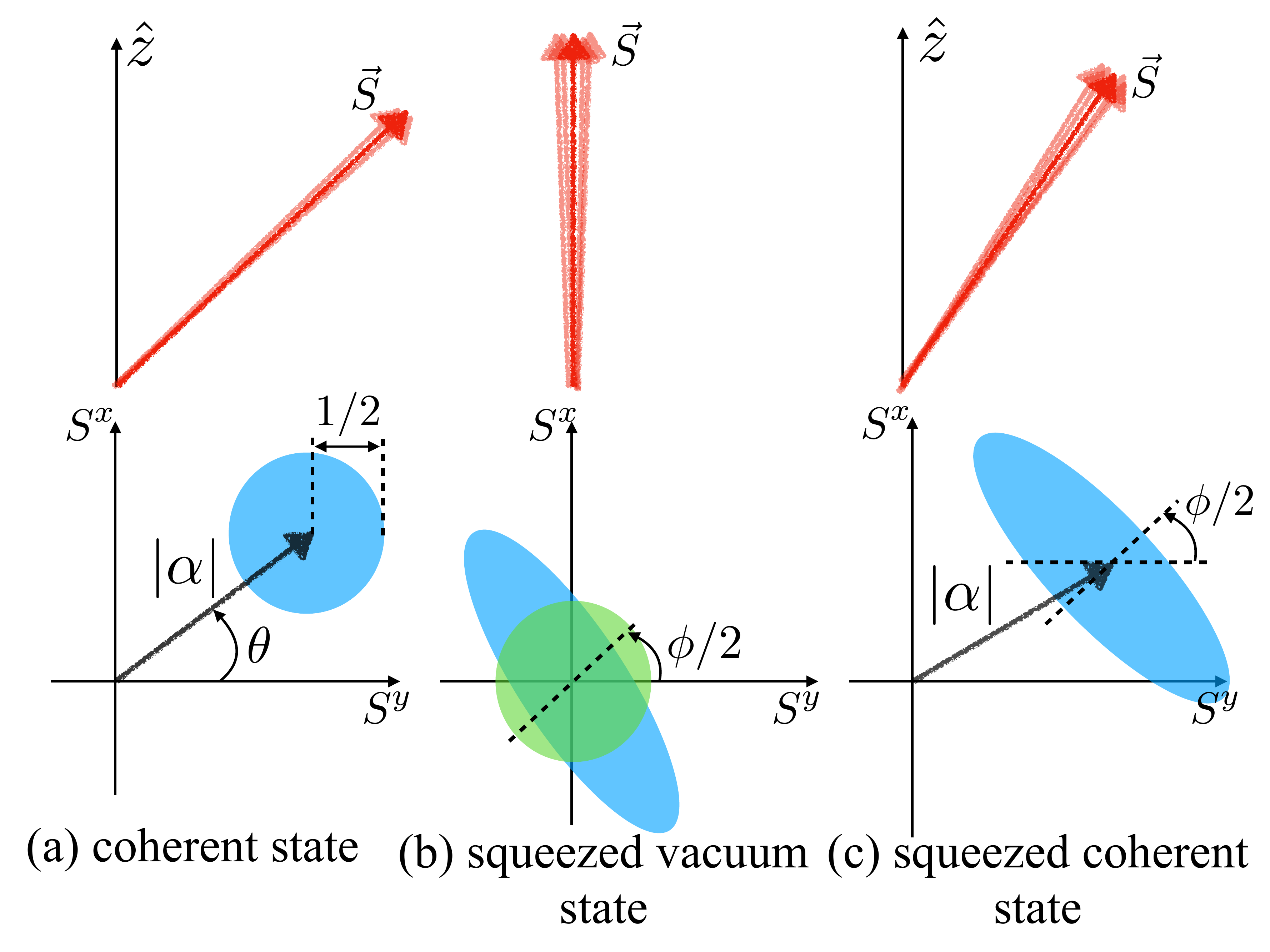}
\caption{(a). A general coherent state $\ket{\alpha}=D(\alpha)\ket{0}$, where $D(\alpha)\equiv \exp{\alpha a^\dagger-\alpha^* a}$ is a displacement operator with $\alpha=|\alpha|e^{i\theta}$,  is the minimum uncertainty state ($\Delta S^x=\Delta S^y=1/2$ ). All spins deviate from $\vb{B}\propto \hat{\vb{z}}$ direction  coherently. (b). The general squeezed vacuum state $S(\zeta)\ket{0}$, where $S(\zeta)\equiv e^{\big[\zeta^*a^2-\zeta(a^\dagger)^2\big]/2}$ is the squeezing operator   with a squeezing parameter $\zeta=re^{i\phi}$, is also a minimum uncertainty state with uncertainties being squeezed (blue ellipse) compared with the vacuum uncertainties (green disk). The direction of the squeezing (the orientation of the semi-minor axis of the ellipse with respect to the $S^x$ axis) is $\phi/2$. The length of the semi-minor axis is $e^{-r}/2$ and the length of the semi-major axis is $e^r/2$. The average direction of the spin in such state is along  $\vb{B}$. (c).  For the general squeezed coherent state $D(\alpha)S(\zeta)\ket{0}$, the degree of the deviation from $\hat{\vb{z}}$ is determined by the parameter $\alpha$ and the degree of squeezing of the uncertainty is determined by the squeezing parameter $\zeta$.}
\label{fig1}
\end{figure}

Staying in the large-$J$ limit, when we turn off both the anisotropy $K$ and the in-plane magnetic field $h$ so that $r=\alpha=0$, the ground state is the Fock vacuum of operators $a_{\vb{k}}$, corresponding to all spins aligned along  $\vb{B}$. If we turn on the in-plane magnetic field $\vb{h}$ and keep the anisotropy off, there are finite number of magnons in the $q=0$ mode forming a coherent state $\ket{\psi}=D(\alpha)\ket{0}$, where all spins deviate from $\vb{B}$ direction uniformly (see Fig. \ref{fig1}a). The number of magnons  (the degree of the deviation) is determined by the magnitude of $\vb{h}$ via $N_c=N_0h^2/4B^2$, which is much smaller than $N_0$ in the dilute limit. We emphasize that this is a  minimum uncertainty state and equally balanced between $S^x$ and $S^y$ with $\Delta S^x=\Delta S^y=1/2$ \cite{uncertainty}. If we turn on the anisotropy and keep the in-plane magnetic field off, the ground state is a squeezed vacuum state $\ket{\psi}=S(r)\ket{0}$, where spins align along $\vb{B}$ on average by noting that $\langle S^x\rangle =\langle S^y\rangle=0$ but with finite number of condensed magnons $N_s=\sinh^2r$ (see Fig.~\ref{fig1}b). The uncertainty is also minimized in this state, but not  equally balanced between $S^x$ and $S^y$. This is also implied from the Hamiltonian $H_1$ where the in-plane $\text{U}(1)$ symmetry is broken explicitly, which is the crucial ingredient for the presence of entanglement. The degree of this squeezing is measured by the squeezing factor $r$, more explicitly $\Delta S^y/\Delta S^x=e^{2r}$. When both anisotroy and in-plane magnetic field are present, all spins deviate from $\vb{B}$ on average and the uncertainties in $S^x$ and $S^y$ have the same behavior as the squeezed vacuum state. The average number of condensed magnons is $N=N_s+N_c$, consisting of coherent magnons and squeezed magnons [see Eq.~(\ref{number})]. The entanglement between two arbitrary spins is determined by the interplay between those parts. In other words, we can tune the entanglement by varying the parameters, such as the anisotropy and the in-plane magnetic field which determine $N_c$ and $N_s$. Note that we refer to this global tilting state as a condensed state, since the number of magnons in the q=0 mode remains to be finite under any spin rotation in the spin space.

In the thermodynamic limit, the contribution to the entanglement due to $q=0$ mode is negligible and the effective ground state is  the squeezed vacuum state $\ket{\psi}=\prod_{\vb{k}\neq 0} S(\phi_{\vb{k}}/2)\ket{0}$. There are only squeezed magnons, because the uniform magnetic field only couples with the $q=0$ mode. The entanglement in this case is distance dependent, since $\ket{\psi}$ involves finite-wavenumber modes. This will be addressed in Sec.~\ref{cc}. Before delving into that, let us introduce the entanglement measure we employ in our analysis: the concurrence. 

\section{Entanglement Measures}
\label{measure}
The problem of measuring entanglement is a vast field of research on its own \cite{RevModPhys.80.517,Plbnio:2007:IEM:2011706.2011707}. Numerous different methods have been proposed to that end. For a pure bipartite state $\rho_{\text{AB}}= \ket{\psi_{\text{AB}}}\bra{\psi_{\text{AB}}}$, we usually adopt the von Neumann entropy as the entanglement measure: $S(\ket{\psi_{\text{AB}}})\equiv -\tr\rho_A\ln \rho_A=-\tr\rho_B\ln \rho_B$.  For a general mixed state $\rho_{\text{AB}}$, this von-Neumann entropy is no longer a good measure since the classical mixture in $\rho_{\text{AB}}$ will also contribute to  the von Neumann entropy. Therefore, many new measures have been introduced, such as entanglement of formation, distillable entanglement, and entanglement cost, which all reduce to the von Neumann entropy when evaluated on pure states. In this paper, we will use the entanglement of formation as the entanglement measure as we can accomplish some analytic results for problems we are interested in. 

The entanglement of formation is defined as
\( E_F(\rho_{\text{AB}}) \equiv \text{min} \sum_i p_i\, S(\ket{\psi^i_{\text{AB}}}),   \)
where the minimum is taken over all possible decompositions of $\rho_{\text{AB}}=\sum_i p_i \ket{\psi^i_{\text{AB}}}\bra{\psi^i_{\text{AB}}}$ and $S(\ket{\psi^i_{\text{AB}}})$ is the von Neumann entropy of the pure state $\ket{\psi^i_{\text{AB}}}$. Physically, $E_F(\rho_{\text{AB}})$ is the minimum amount of pure state entanglement needed to create the mixed state. This is extremely difficult to evaluate in general since we need to try all the decompositions. Quite remarkably an explicit expression of $E_F(\rho_{\text{AB}})$ is given when both $A$ and $B$ are two-state systems (qubits). This exact formula is based on the often used two-qubit concurrence, which is defined as \cite{PhysRevLett.80.2245}
\( \mathcal{C}(\rho)=\text{max} \{ 0,\lambda_1-\lambda_2-\lambda_3-\lambda_4 \}, \label{concurrence} \)
where $\lambda_i$'s are, in decreasing order, the square roots of the eigenvalues of the matrix $\rho (\sigma_y\otimes \sigma_y)\rho^*(\sigma_y\otimes \sigma_y)$, where $\rho^*$ is the complex conjugate of $\rho$. The entanglement of formation is then given by
\begin{eqnarray}
E_F(\rho)&=& h\Big( \frac{1+\sqrt{1-\mathcal{C}^2}}{2}  \Big),  \\
h(x)&=&-x\log_2x-(1-x)\log_2(1-x).
\end{eqnarray}
$E_F(\rho)$ is monotonically increasing and ranges from 0 to 1 as $\mathcal{C}(\rho)$ goes from 0 to 1, so that one can take the concurrence as a measure of entanglement in its own right. 
We include two examples in Appendix \ref{appendix:a} and we will use the result of the second example in our following analysis. In the next section, we explore the entanglement between two arbitrary spins for various states we discussed in Section \ref{model}.

\section{Entanglement Quantification}
\label{main}
In the first four subsections, we discuss the entanglement due to the $q=0$ mode in a finite size sample with large exchange coupling where this mode dominates the quantum fluctuation. In the last subsection,  entanglement and its distance dependence are examined in thermodynamic limit, where finite wavenumber modes must be taken into account.

\subsection{Fock States}
We start with investigating the concurrence between two spins in Fock states $\ket{N}$  where $N$ is the number of magnons in the zero-momentum mode. For the Fock vacuum $\ket{0}$, the concurrence is zero since this is a product state $\ket{\uparrow\uparrow\cdots \uparrow}$. When there is a finite number of magnons,  invoking the reduced density matrix (\ref{density matrix}), which is simplified for this specific case, we show the concurrence (Eq. \ref{concurrence}) between arbitrary two spins is given by
\begin{eqnarray}
\mathcal{C}_{\text{Fock}}&=& 2 \,\text{max}\bigg\{   0,   | \langle   \sigma_i^+\sigma_j^-  \rangle |  - \sqrt{  \langle  k_i^+k_j^+    \rangle  \langle  k_i^-k_j^-   \rangle   }  \, \bigg \} \nonumber \\
&\approx& 2\frac{N}{N_0}\big(1-\sqrt{1-1/N}\,  \label{fock}  \big).
\end{eqnarray}
Note that the upper bound of the concurrence is $2/N_0$, which is known as the tight bound for symmetric sharing of entanglement \cite{PhysRevA.62.050302}. The concurrence  reaches its maximum value when there is only one magnon, corresponding to the state $\ket{\downarrow\uparrow\cdots \uparrow}+\ket{\uparrow\downarrow\cdots \uparrow}+\ket{\uparrow\uparrow\cdots \downarrow}$. This is a generalization of the Bell state $\ket{\Psi^+}\sim \ket{\uparrow\downarrow}+\ket{\downarrow\uparrow}$ and thus maximally entangled. Another feature we should pay attention to is that the concurrence is a decreasing function of $N$ and approaches $1/N_0$ as $N\rightarrow \infty$. This is consistent with the analysis of the Dicke state \cite{Friedberg_2007} $\ket{N_0/2,M}$ in quantum optics, which describes a system consisting of $N_0$ two-level systems (spin-1/2 particles)  and is a pure symmetric (with respect to permutations) state. $N_0/2-M$ is the number of excited two-level systems (i.e., the flipped spins). The  concurrence of such Dicke state is given by \cite{MA201189,PhysRevA.68.033821,Wang2002}
\begin{eqnarray}
\mathcal{C}_{\text{Dicke}}&=&\frac{N_0^2-4M^2-\sqrt{ (  N_0^2-4M^2  )[(N_0-2)^2-4M^2  ]   }}{2N_0(N_0-1)}\nonumber \\
&\approx& \mathcal{C}_{\text{Fock}},
\end{eqnarray}
where we have identified $N=N_0/2-M$ and specialized to the case $N\ll N_0$ by noting that the number of excited two-level systems is exactly the number of magnons in our context. It should be clear from our discussion above that we must invoke the one-magnon state $\ket{N=1}$ to produce a maximally-entangled configuration.

In Ref.~\cite{PhysRevB.97.060405}, it was found that the entanglement between two spins increases with the number of condensed magnons $N$, which is contrary to what we discussed above. This discrepancy can be traced to the second term in Eq.~(\ref{fock}), which, despite being comparable to the first term, was omitted in Ref.~\cite{PhysRevB.97.060405}. In particular, we see that the entanglement vanishes in the thermodynamic limit, $N_0\to\infty$, in the Fock state $\ket{N}$ with any $N$, in agreement with the tight bound for symmetric sharing of entanglement. At a finite temperature $T$, when two spins sit at a distance smaller than the thermal de Broglie wavelength  $\lambda_T\propto \sqrt{J/T}$, we expect the concurrence to be inversely proportional to the total number of sites within the corresponding volume: $\mathcal{C}\propto 1/\lambda_T^3$ (crossing over to $\mathcal{C}\propto 1/N_0$ as $T\rightarrow 0$ and $\lambda_T$ exceeds the system size). Beyond the thermal de Broglie wavelength  $\lambda_T$, the entanglement should decay exponentially, $\sim e^{-R/\lambda_T}$, with the distance $R$ between two spins, which agrees qualitatively with the analysis of Ref.~\cite{PhysRevB.97.060405}.

\subsection{Coherent States}
Let us turn on the in-plane magnetic field which will lead to  a coherent state $\ket{\alpha}\equiv D(\alpha)\ket{0}$ as the ground state. We evaluate the elements of density matrix (\ref{density matrix})  in the coherent state and obtain \( \rho (\sigma_y\otimes \sigma_y)\rho^*(\sigma_y\otimes \sigma_y)\propto 1_{4\times 4}.\) This implies the entanglement between any two spins is zero and there is no  quantum correlation stored in spins according to the definition of the concurrence Eq. (\ref{concurrence}). Indeed, this is not surprising and it has been shown any bosonic coherent state is unentangled \cite{PhysRevA.66.052323}. It is true as well for spin coherent states since $\ket{\theta,\phi}=\otimes_{l=1}^{N_s}\big[ \cos\frac{\theta}{2}\ket{\downarrow}_l +e^{i\phi}\sin\frac{\theta}{2} \ket{\uparrow}_l   \big] $ is a  product state where $\theta$ and $\phi$ specify the direction of spins.  Coherent states have minimum uncertainty which are equally balanced between $S^x$ and $S^y$ with $\Delta S^x=\Delta S^y=1/2$.   Furthermore, any classical mixture of coherent states, such as $\hat{\rho}=\int d^2\alpha\,\,  P_\alpha \ket{\alpha}\bra{\alpha}$ with $P_\alpha>0$ being the probability density in $\ket{\alpha}$,  can only increase the uncertainty and also has zero entanglement since classical correlations do not contribute to the entanglement. Such  states are known as classical light states in the quantum optics \cite{Sanders_2012}. One typical nonclassical light state is squeezed states and we will examine the entanglement of those states below.

\subsection{Squeezed Vacuum Magnetic States}
By turning on the anisotropy in Hamiltonian $H_1$ and keeping the in-plane magnetic field off, we can generate the squeezed vacuum state $\ket{\psi}=S(r)\ket{0}$ as the ground state, where the uncertainty in $S^x$ is below the vacuum level. There must be quantum correlations in such states since they can never be achieved by mixing coherent states. We show that the concurrence (\ref{concurrence})  between two spins is given by
\( \mathcal{C}=\frac{2}{N_0} \frac{\sqrt{N_s}}{ \sqrt{N_s+1} +\sqrt{N_s   } },   \label{squeeze vacuum}\)
where $N_s=\sinh^2r$ is the number of magnons in the squeezed vacuum state. In contrast to the Fock state, the concurrence of the squeezed vacuum state increases as we increase the number of magnons. This can be understood by noting that increasing $N_s$ corresponds to squeezing the vacuum more. Namely, the degree of squeezing, $\Delta S^y/\Delta S^x=e^{2r}=(\sqrt{N_s+1}+\sqrt{N_s} )^2  $, equals unity when $N_s=0$, which corresponds to zero entanglement. $\Delta S^y/\Delta S^x$  approaches infinity as $N_s$ rises, where the vacuum is infinitely squeezed and has maximum concurrence $1/N_0$, which is half of the tight bound for symmetric sharing of entanglement \cite{PhysRevA.62.050302}. We remark that this reduction in the entanglement is due to the odd parity missing in the wave function.

\subsection{Squeezed Coherent Magnetic States}
\begin{figure}
\includegraphics[scale=0.33]{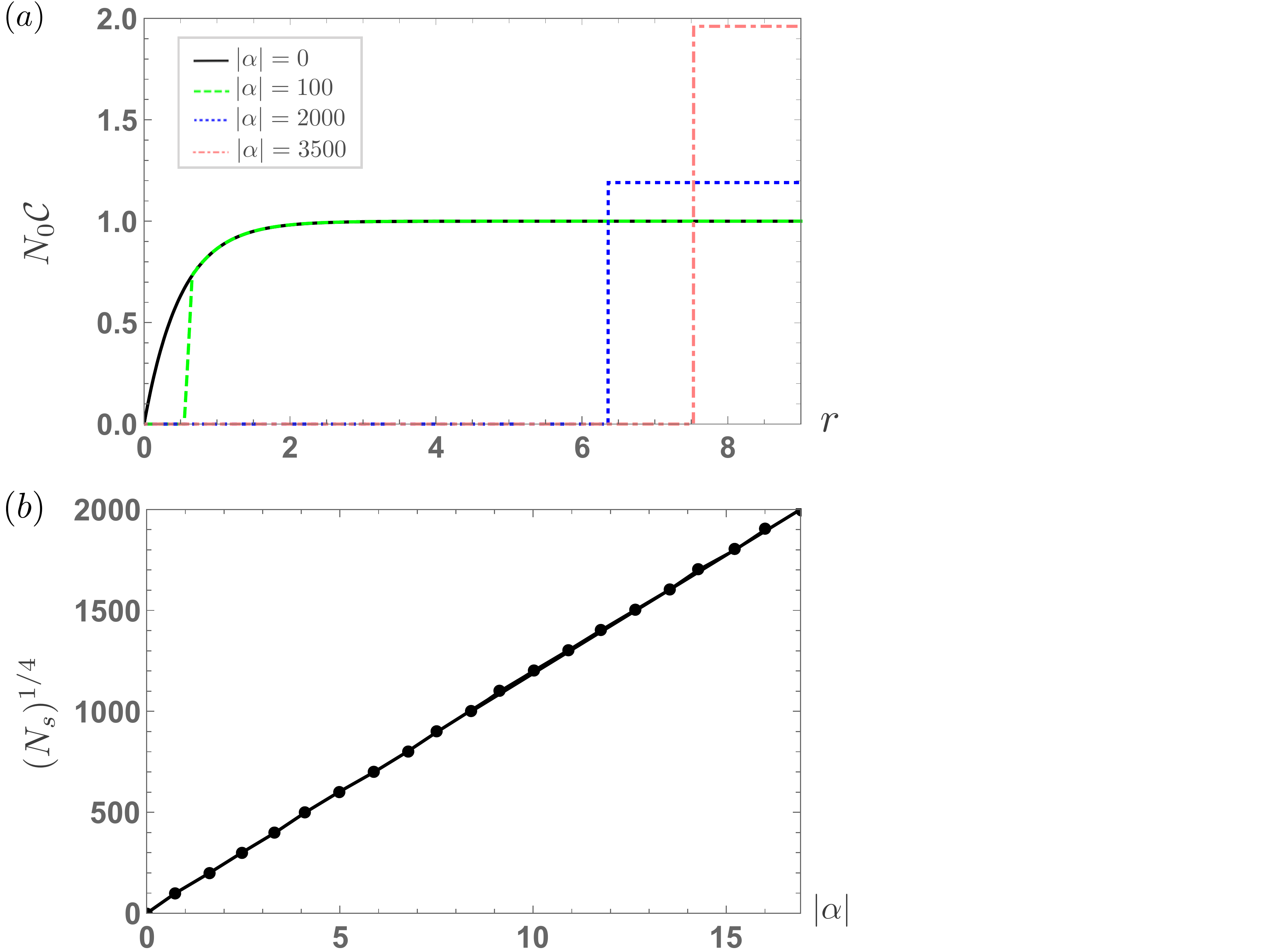}
\caption{(a). Concurrence $\mathcal{C}$ as a function of the number of squeezed magnons $N_s=\sinh^2r$ for different number of coherent magnons $N_c=|\alpha|^2$.  $|\alpha|=0$ corresponds to the squeezed vacuum state (see Eq. (\ref{squeeze vacuum})), where the maximal concurrence  is $1/N_0$. As we increase the number of coherent magnons, the physics is dominated by the coherent part when the number of squeezed magnons is small. As a result, the concurrence is zero. However, the presence of the coherent part can increase the upper bound of the concurrence once the number of squeezed magnons is above some critical value $\sinh^2r_c$, which depends on the amount of coherent magnons we put into the system.  (b). Critical values $(N_s)^{1/4}=\sqrt{\sinh r_c}$ as a function of $|\alpha|$, which can be fitted using a linear relation. When the number of coherent magnons is larger than $\sqrt{2N_0 N_s}$, the coherent part dominates the physics and the concurrence is zero. Otherwise, we have a finite concurrence. In the numerical study above, we set $N_0=10^8$.}
\label{fig2}
\end{figure}

Now let us turn on both the anisotropy and the in-plane magnetic field, resulting in the squeezed coherent state $\ket{\psi}=D(\alpha)S(r)\ket{0}$ as our ground state. We will see that in contrast to the coherent states that retain their (trivial) entanglement character under displacement, displacing a squeezed state does have a nontrivial effect. Unlike states we discussed above, however, it is difficult to obtain an analytic expression for the concurrence in this case. Therefore, we obtain the concurrence $\mathcal{C}$ numerically by  plotting it  as a function of $r$ for different values of $|\alpha|$ (see Fig. \ref{fig2}a), where we assume the anisotropy $K>0$ without loss of generality. As increasing the value of $|\alpha|$ from zero, we have zero concurrence under a critical value of $r_c=r_c(|\alpha|)$ (see Fig. \ref{fig2}b) and  nonzero concurrence above $r_c$.  The maximal concurrence will increase to $2/N_0$ from $1/N_0$ as we increase the number of coherent magnons. Thus we can see that the coherent magnons will unlink the quantum correlations between spins established by a small number of squeezed magnons. This is because the coherent magnons dominate the physics when the number of squeezed magnons is small. However, coherent magnons will be beneficial for information storage when the number of squeezed magnons is large. The upper bound of the concurrence rises since the coherent part  involves states with both parities, unlike squeezed vacuum states which only involve  even parity states $\{\ket{2k}\}$, and thus increases the upper bound. We remark that, in contrast to the discussion in Sec. \ref{main} B where the displacement operation does not yield entanglement since it is acting on a trivial state (Fock vacuum state), the displacement operator here results in nontrivial entanglement behavior as it acts on a squeezed state which is entangled.

To determine this transition,  we study the critical value $N_s^{1/4}=\sqrt{\sinh r_c}$ numerically and plot it as a function of $|\alpha|$ which can be fitted well with a linear relation (see Fig. \ref{fig2}b). We conclude that there is no entanglement when
\( N_c\geq \sqrt{2N_0 N_s} \label{transition}. \)
Otherwise we have nonzero entanglement. This transition is discontinuous as implied from Fig. \ref{fig2}a. When $|\alpha|$ is large, $N_0\mathcal{C}$ is a step function of $r$, which can be potentially used as an efficient switch in quantum information processing tasks.

\subsection{Thermodynamic Limit}
\label{cc}
\begin{figure}
\includegraphics[scale=0.35]{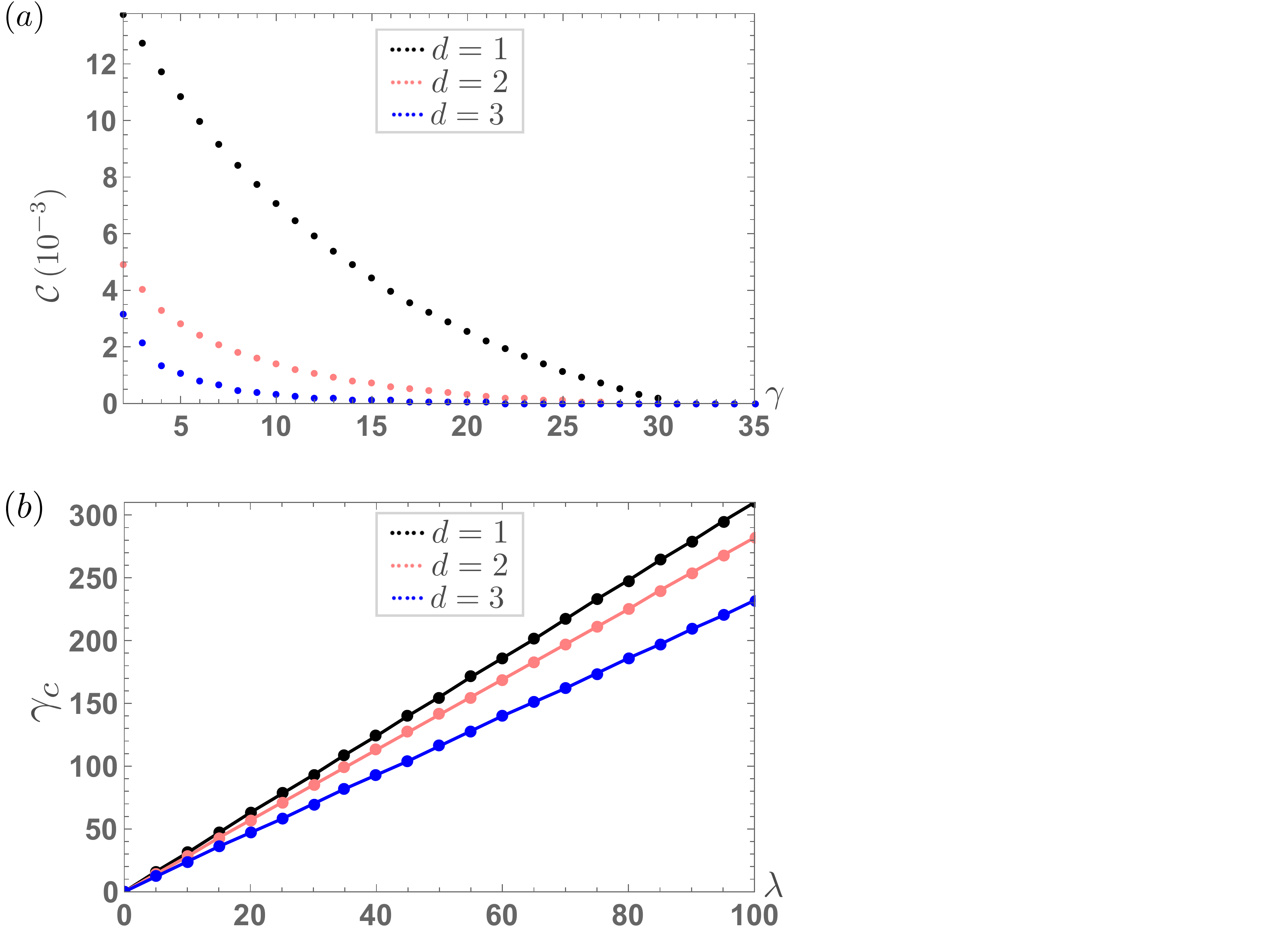}
\caption{(a). Concurrence as a function of the distance $\gamma$ between two spins in dimension  $d=1$, $d=2$ and $d=3$, respectively. The overall value of the concurrence will decrease as we increase the dimension of our system. In a given dimension, the concurrence is finite but will decrease as we increase the distance between these two spins within a critical distance $\gamma_c$, beyond which the concurrence vanishes.  We have set $\lambda=10$, $\eta=0.5$ and $\hat{\vb{R}}=\hat{\vb{x}}$ in higher dimension. (b). Critical distance $\gamma_c$ as a function of the correlation length $\lambda$, which can be fitted well with a linear relation. We have set $\eta=0.5$.}
\label{fig4}
\end{figure}

In the thermodynamic limit, one can see that the entanglement between two arbitrary spins due to $q=0$ mode vanishes from our discussion above. Under the circumstances, nonzero wavenumber modes should be taken into account and we can show that the concurrence between a spin at $\vb{R}_i$ and a spin at $\vb{R}_j$ in $d$ dimension is given by 
\(\mathcal{C}_{ij}=\max\big\{0, \mathcal{T}(\gamma, \lambda, \eta)\big\}, \label{thermo} \)
where
\begin{multline}
\mathcal{T}(\gamma, \lambda, \eta) \approx      \frac{1}{(2\pi)^d}    \bigg|        \int_{\text{B.Z.}} d^d\vb{q}  \frac{\eta \cos(\gamma \hat{\vb{R}}\cdot \vb{q} )  }{\sqrt{  (1+ 2\lambda^2\vb{q}^2  )^2   -\eta^2     }}     \bigg|    \\
 \,\,\,\,\,+   \frac{1}{(2\pi)^d}    \int_{\text{B.Z.}}  d^d\vb{q} \bigg[ 1  -\frac{ 1+ 2\lambda^2\vb{q}^2    }{\sqrt{  (1+ 2\lambda^2\vb{q}^2  )^2   -\eta^2     } }  \bigg]. 
\end{multline}
The derivation is given in Appendix \ref{thermoc}. Here $\gamma=|\vb{R}|=|\vb{R}_i-\vb{R}_j|$ and $\lambda=\sqrt{J/B}$  are the distance between two spins and the exchange length. $\eta=|K|/B<1$ is a dimensionless parameter and $\hat{\vb{R}}=(\vb{R}_i-\vb{R}_j)/|\vb{R}|$. $\text{B.Z.}$ represents the Brillouin zone.  Figure \ref{fig4}a visualizes the distance $\gamma$ dependence of concurrence $\mathcal{C}_{ij}$ in dimension $d=1$, $d=2$ and $d=3$, respectively. It suggests that $\mathcal{C}_{ij}$ is smaller when the dimension is higher. From Fig. \ref{fig4}b, we can see that, keeping other parameters fixed, $\mathcal{C}_{ij}$ will decrease to zero as we increase the distance $\gamma$ to a critical value $\gamma_c$, which is proportional to the exchange length $\lambda$. In other words, spins within the exchange length can communicate and entangle with each other. Note that, in the limit of diverging exchange length $\lambda\rightarrow \infty$, the overall value of concurrence will vanish even though spins can entangle with each other over a long distance. 

\subsection{Remarks}

Let us modify the $H_1$ and $H_2$ to allow for more general squeezed coherent states,
\begin{eqnarray}
H_1=& \frac{K}{w} \sum_{\langle ij \rangle}   \big[        \cos\theta_1 (  S^x_iS^x_j-S^y_iS^y_j     ) \nonumber \\
&\,\,\,\,\,\,\,\,\,+2\sin\theta_1 S^x_iS^y_j \big],\\
\nonumber\\
H_2=&-h \cos\theta_2 \sum_i S^x_i -h \sin\theta_2 \sum_i S^y_i.
\end{eqnarray}
Compared with  the original Hamiltonian (\ref{totalH}), we have rotated the in-plane magnetic field and anisotropy by $\theta_2$ and $\theta_1/2$, respectively, $H_1\rightarrow U(\theta_1/2)H_1 U(\theta_1/2)^\dagger$ and $H_2\rightarrow U(\theta_2)H_2 U(\theta_2)^\dagger$ with $U(\theta)=\prod_{i}e^{-i\theta S_i^z}$ being the rotation operator. Therefore, the entanglement should only depend on the physical angle $\theta_2-\theta_1/2$. The ground state of this Hamiltonian, in the large exchange-coupling limit (so that we can neglect nonzero wavenumber modes \cite{nonzero}), is given by $\ket{\psi}=D(\alpha)S(\zeta)\ket{0}$ with $\zeta=re^{i\theta_1}$ and $\alpha=\sqrt{N_0}h[e^{i\theta_2}\cosh 2r -e^{i(\theta_1-\theta_2)}\sinh 2r ]/2\omega$, where $\omega=\sqrt{B^2-K^2}$ and $r$ is determined by $\tanh 2r=K/B$. We recover what we have obtained before [see Eq. (\ref{state})] when $\theta_1=\theta_2=0$, as expected. We point out that, by taking this angle dependence into account, the behavior of the concurrence does not get modified qualitatively, since its angular variation is much smaller than the absolute value (see Fig.~\ref{fig3}).

The above Hamiltonian realizes the general squeezed coherent state $D(\alpha)S(\zeta)\ket{0}$ with $\alpha=|\alpha|e^{ i\theta}$ and $\zeta=re^{i\phi}$ being complex-valued, where $D(\alpha)=e^{\alpha a^\dagger-\alpha^* a}$ is the displacement operator and $S(\zeta)=e^{[\zeta^* a^2-\zeta (a^\dagger)^2]/2}$ is the squeezing operator. The entanglement in such states only depends on $|\phi/2-\theta|$ instead of depending on these two angles separately. This is implied from the Fig.~\ref{fig1}c where the only physical angle is  $|\phi/2-\theta|$. More explicitly, we have $D(|\alpha|e^{i\theta})S(re^{i\phi})\ket{0}\rightarrow D(|\alpha|e^{i(\theta-\phi/2 ) })S(r)\ket{0}$ under a  gauge transformation $a\rightarrow a e^{i\phi/2}$ which will not alter any physics of the system. We find numerically the concurrence is periodic with period $2\pi$ in $\phi$ (the pink curve in Fig.~\ref{fig3}) and $\pi$ in $\theta$ (the black curve in Fig.~\ref{fig3}). This is consistent with what we discussed above where $\zeta=re^{i\theta_1}$ and $\alpha=\sqrt{N_0}h[e^{i\theta_2}\cosh 2r -e^{i(\theta_1-\theta_2)}\sinh 2r ]/2\omega$. Under the  gauge transformation, we have the ground state $D(\alpha e^{-i\theta_1/2})S(r)\ket{0}$
 where $\alpha e^{-i\theta_1/2}=\sqrt{N_0}h[e^{i(\theta_2-\theta_1/2)}\cosh 2r -e^{-i(\theta_2-\theta_1/2)}\sinh 2r ]/2\omega $ and thus the entanglement only depends on the physical angle $\theta_2-\theta_1/2$.
 
\begin{figure}
\includegraphics[scale=0.24]{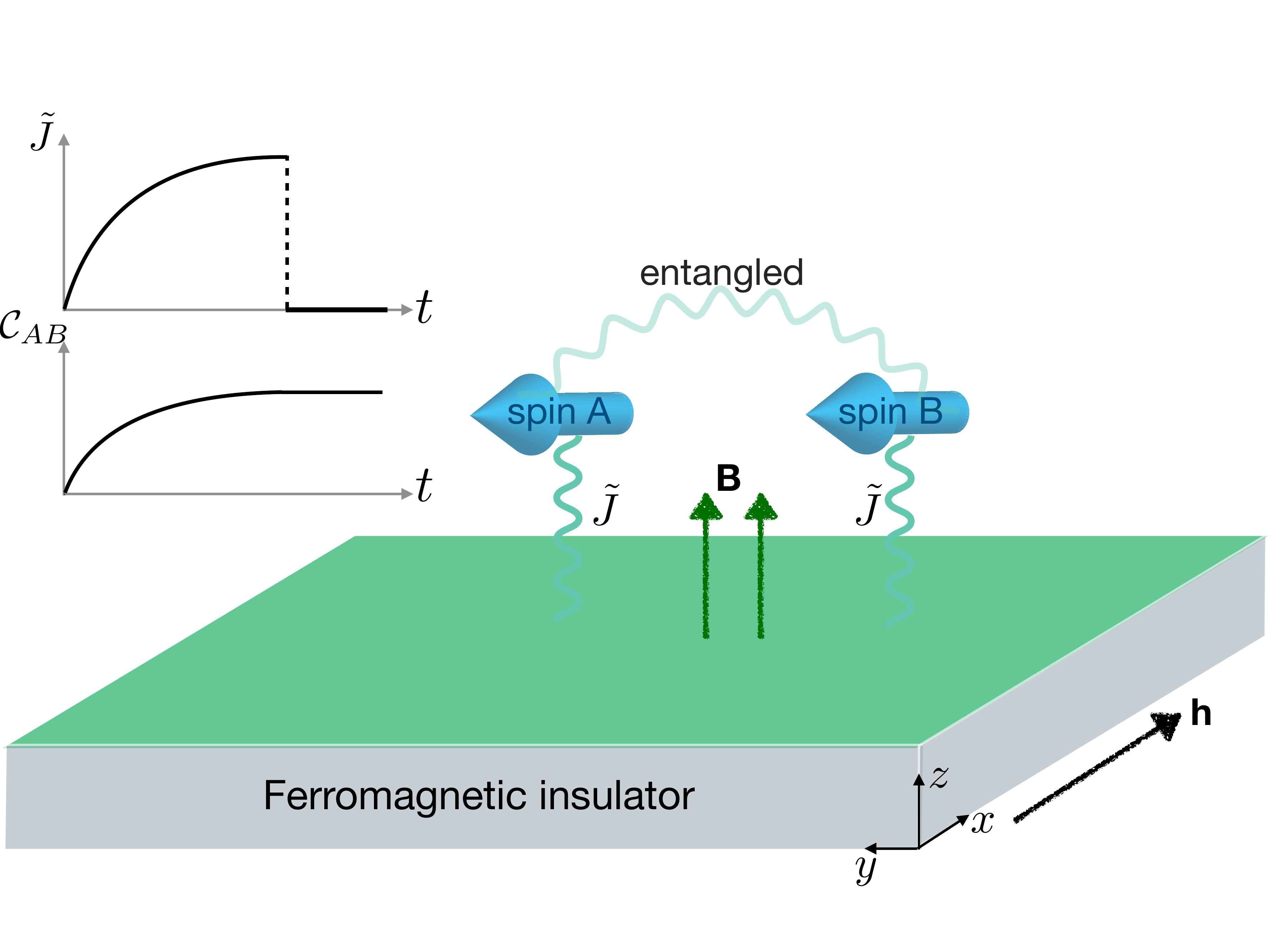}
\caption{Spin A and B are placed above a ferromagnetic insulator subjected to magnetic fields $\vb{h}$, $\vb{B}$ and anisotropies, which realizes  the Hamiltonian that we discussed in Sec. \ref{model}.  Turning on the coupling $\tilde{J}(t)$ between spins and the insulator adiabatically so that these spins behavior like a part of the insulator and thus the concurrence $\mathcal{C}_{AB}$ grows correspondingly. This entanglement remains even after the coupling $\tilde{J}$ is turned off so long as this turnoff process is rapid enough.}
\label{fig5}
\end{figure}

\begin{figure}
\includegraphics[scale=0.24]{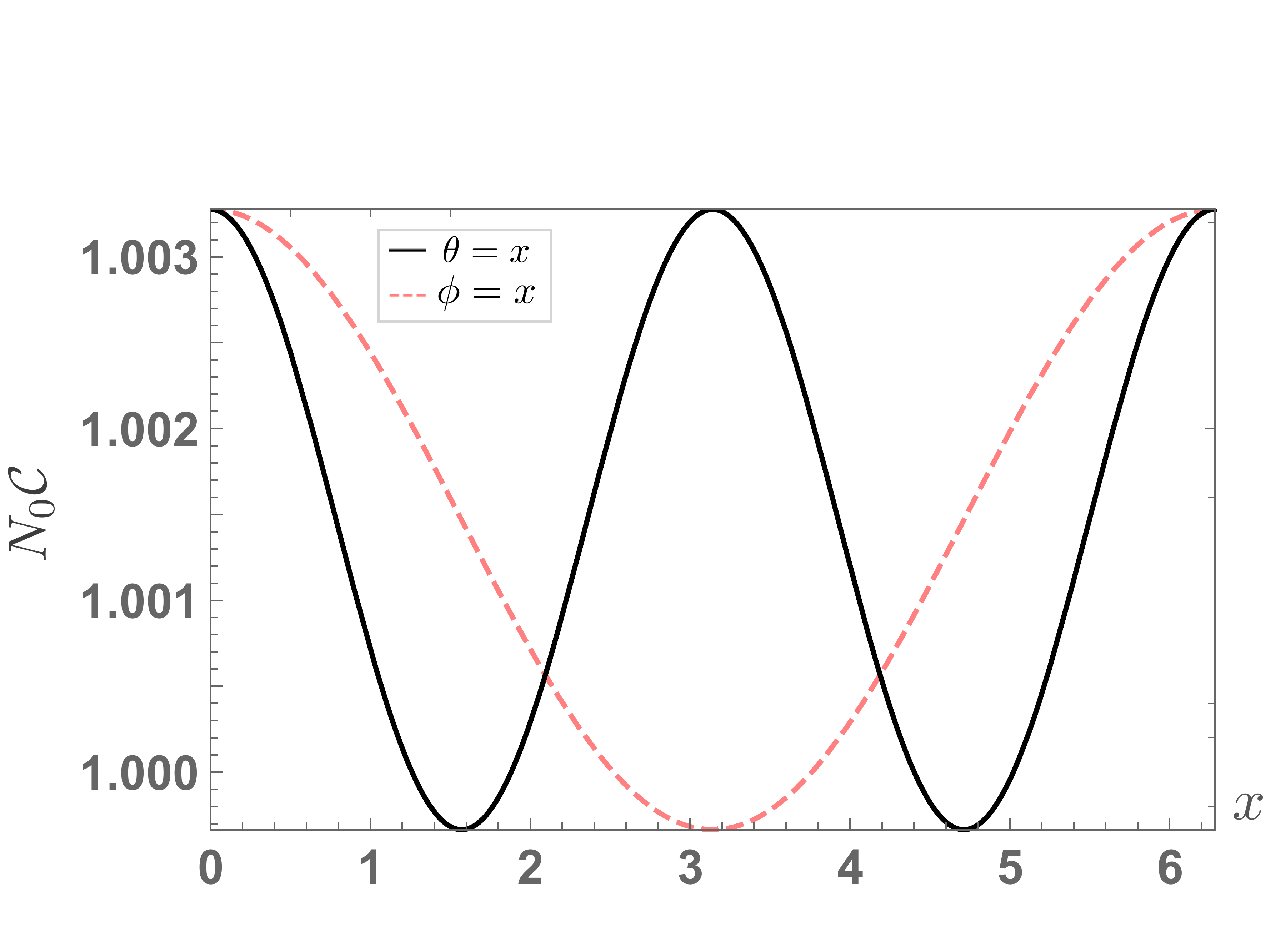}
\caption{Concurrence as a function of one angle (setting the another angle to zero). For the pink dashed  curve,  we plot the concurrence as a function of the angle $\phi$ and set $\theta=0$. For the black solid curve, we plot the concurrence as a function of the angle $\theta$ and set $\phi=0$. We find that concurrence is periodic with period $2\pi$ in $\phi$ and $\pi$ in $\theta$.  In both cases, we set $N_s=\sinh^24, N_c=9\times10^4$ and $N_0=10^8$. }
\label{fig3}
\end{figure}

A few ways have recently been proposed to store and control quantum information in magnetic systems. For example,  	topological defects can be used as quantum information carriers \cite{PhysRevB.97.064401} and two spins can be coupled via the spin-superfluid mode harbored by an antiferromagnetic domain wall \cite{PhysRevB.99.140403}. Our discussion can be also applied to entangle two distant spin qubits (see Fig. \ref{fig5}). Let us consider the situation where two spin qubits are placed in the vicinity of a ferromagnetic insulator. Then, turn on the coupling $\tilde{J}$ between spin qubits and the insulator strong enough so that the two spins are entangled as if they are part of the magnetic insulator. Upon the sufficiently rapid turnoff of the coupling, we can obtain the isolated system of the two spin qubits that remain entangled.

\section{Summary and Outlook}
\label{summary}
The purpose of this paper has been to investigate the entanglement generation and entanglement control in a magnetic system. In the low temperature regime $T\ll J$, magnons  form a general squeezed coherent state, which is a minimum uncertainty state with the quantum noise in one observable reduced below its vacuum level  with the sacrifice of enhanced uncertainties in the another observable. We showed these squeezed states can be fully controlled by tuning applied external fields and in-plane anisotropies.  Utilizing the entanglement of formation, or more specifically the concurrence, as a measure of entanglement, we illustrated that in the large exchange-coupling limit, a general squeezed coherent state, including its special case of a squeezed vacuum state, exhibits a high  degree of entanglement between two arbitrary spins, as opposed to a coherent state which is not entangled.   Therefore, a magnetic system can serve as a resource for storing quantum information and processing quantum information, such as quantum teleportation, quantum network and quantum logical encoding.

As temperature rises, we expect a thermal crossover from squeezing dominated regime to simply Fock-coherent regime discussed in Sec. \ref{main}A. A more systematic study of the temperature dependence of entanglement is left for a future work. In our analysis, we ignored the dipole-dipole interaction. For the uniform mode, the dipolar interactions would simply contribute a shape-dependent demagnetizing field, which can be absorbed into our anisotropy constants \cite{kittel}. For the large-$\vb{k}$ modes, the effective anisotropies would become $\vb{k}$-dependent, which would modify the quantitative details of our analysis. At this point, for simplicity, we are focusing on the materials where dominant anisotropies are crystalline. Two-mode squeezing arises naturally also in Heisenberg antiferromagnets \cite{kamra}, resulting in a large entanglement between two antiparallel magnetic sublattices even in the absence of magnetic anisotropies. It may be interesting to study the spatial distribution of this entanglement as well as its possible tunability by external parameters.

In the thermodynamic limit, the entanglement attributable to the zero wavenumber mode vanishes due to the existence of the tight bound for symmetric sharing of entanglement $2/N_0$. Thus nonzero wavenumber modes should be taken into account and we studied the distance dependence of the concurrence.   The existence of the tight bound is because  we are considering the entanglement between two spins.  How will the entanglement bound change   if we consider the entanglement between two regions (that can contain many spins separately in general) instead of just two spins? This scaling property of the entanglement is well understood when the bipartite system is a gapped ground state of a local Hamiltonian  and known as the entanglement area law \cite{RevModPhys.82.277}. Its constant correction is known as the topological entanglement entropy \cite{PhysRevLett.96.110404}  characterizing many-body states that possess topological order. For a  mixed state, however, as in our case, the scaling property is far from being well understood. Nevertheless, we would expect  the upper bound of the entanglement to increase as  we consider the entanglement between two regions in general, since the Hilbert space is larger compared with the two-spin case. Therefore, it can potentially store more quantum information, with the exact scaling behavior remaining to be explored.


\begin{acknowledgments}
This work was supported in part by NSF under Grant No.~DMR-1742928 (J.Z. and Y.T.) and by Research Council Grant URC-19-090 of the University of Missouri (S.K.K.).
\end{acknowledgments}

\appendix
\section{Examples of Concurrence}
\label{appendix:a}
\subsection{Two Qubits}
Before delving into many body states, let us try out the concurrence for a two-qubit system. Assuming the density matrix is given by:
\(\rho=(1-p)\ket{\uparrow\uparrow}\bra{\uparrow\uparrow}+p\ket{\text{singlet}}\bra{\text{singlet}},\)
 with probability $1-p$ in state $\ket{\uparrow\uparrow}$ and $p$ in state $\ket{\text{singlet}}\equiv(\ket{\uparrow\downarrow}-\ket{\downarrow\uparrow} )/\sqrt{2}$. In the basis of $\ket{\uparrow\uparrow},\ket{\uparrow\downarrow},\ket{\downarrow\uparrow},\ket{\downarrow\downarrow}$,
\(\rho=\begin{bmatrix}
1-p & 0 & 0 & 0\\
0 & p/2 & -p/2 & 0\\
0 & -p/2 & p/2 & 0\\
0 & 0 & 0 & 0\\
\end{bmatrix}. \label{a2}\)

We would expect    the concurrence  will increase as we increase  $p$ since $\ket{\uparrow\uparrow}$ is not entangled but  $\ket{\text{singlet}}$ is entangled. We can compute the square roots of the eigenvalues of the matrix $\rho (\sigma_y\otimes \sigma_y)\rho^*(\sigma_y\otimes \sigma_y)$ exactly and we have  $\lambda_1=p,\lambda_2=0,\lambda_3=0,\lambda_4=0$, thus
\(\mathcal{C}(\rho)=\text{max}\{ 0,p \}=p, \)
which is exactly what one might expect.
\subsection{$N$ Qubits}
For a $N$-qubit system, whose dynamics is governed by a Hamiltonian $H$, assuming the system is in a thermal equilibrium, we can calculate the entanglement between two arbitrary qubits $i$ and $j$. The reduced density matrix of those two qubits is obtained by tracing out other degrees of freedom and given by
 \( \rho_{ij}=\frac{1}{4} \sum_{\alpha,\beta}p_{\alpha\beta} \sigma_i^\alpha\otimes \sigma_j^\beta,      \)
where $\sigma^\alpha=\{ I,\sigma^x,\sigma^y,\sigma^z\}$ and $p_{\alpha\beta}=\langle  \sigma_i^\alpha\otimes \sigma_j^\beta \rangle=\tr ( e^{-\beta H}   \sigma_i^\alpha\otimes \sigma_j^\beta) /Z $ is real. $Z$ is the partition function and $\beta=1/k_BT$. In the same basis as Eq. (\ref{a2}),  the explicit form of $\rho_{ij}$ is given by \cite{PhysRevB.97.060405,PhysRevA.66.032110}
\( \rho_{ij} = \mqty[ \langle  k_i^+k_j^+    \rangle     &   \langle  \sigma_i^-k_j^+   \rangle   &  \langle  k_i^+\sigma_j^-   \rangle    &    \langle  \sigma_i^-\sigma_j^-   \rangle       \\     \langle  \sigma_i^+k_j^+   \rangle       &   \langle  k_i^-k_j^+   \rangle   &  \langle   \sigma_i^+\sigma_j^-  \rangle    &    \langle  k_i^-\sigma_j^-   \rangle    \\    \langle  k_i^+\sigma_j^+   \rangle     &  \langle  \sigma_i^-\sigma_j^+   \rangle    &  \langle  k_i^+k_j^-   \rangle    &   \langle  \sigma_i^-k_j^-   \rangle   \\   \langle  \sigma_i^+\sigma_j^+   \rangle   &  \langle  k_i^-\sigma_j^+   \rangle    &    \langle   \sigma_i^+k_j^-  \rangle  &  \langle  k_i^-k_j^-   \rangle  ],   \label{density matrix}    \)
where $k^{\pm}=(1\pm \sigma_z)/2 $, $\sigma^{\pm}=(\sigma^x\pm i\sigma^y)/2$ and we have dropped the tensor product symbol. We are now ready to evaluate the concurrence once the Hamiltonian is specified. 

\section{Concurrence in Thermodynamic Limit}
\label{thermoc}
Here we sketch the derivation of Eq. (\ref{thermo})  (Eq. (\ref{fock}) and Eq. (\ref{squeeze vacuum}) are similar). In state $\ket{\psi}=\prod_{\vb{k}\neq 0} S(\phi_{\vb{k}}/2)\ket{0}$, where $S(\phi_{\vb{k}}/2)=e^{(a_{\vb{k}}a_{-\vb{k}}-a^\dagger_{\vb{k}}a^\dagger_{-\vb{k}})\phi_{\vb{k}} /2}$ is the two-mode squeezing operator  and $\phi_{\vb{k}}$ is determined by $\tanh\phi_{\vb{k}}=K/(2J \vb{k}^2+B)$, we evaluate the reduced density matrix Eq. (\ref{density matrix}) for a spin at  $\vb{R}_i$ and  a spin at $\vb{R}_j$ in $d$ dimension and  obtain
\begin{widetext}
\(\rho_{ij}=   \mqty[ 1-2\langle a^\dagger_i a_i \rangle +\langle a^\dagger_ia_ia_j^\dagger a_j\rangle    &  0   & 0   &    \langle  a^\dagger_i a^\dagger_j  \rangle       \\     0     &  \langle a^\dagger_i a_i \rangle-  \langle a^\dagger_ia_ia_j^\dagger a_j\rangle    &  \langle  a^\dagger_j a_i  \rangle    &    0   \\    0    &  \langle  a^\dagger_i a_j  \rangle    & \langle a^\dagger_j a_j \rangle-  \langle a^\dagger_ia_ia_j^\dagger a_j\rangle    &  0  \\  \langle a_i a_j \rangle   &  0    &   0  &  \langle a^\dagger_ia_ia_j^\dagger a_j\rangle      ].    \)
\end{widetext}
Here we have used the fact that expectation value of any product of odd number of magnon creation or annihilation operators vanishes (this is true for any squeezed vacuum state), for example $\bra{\psi}a^\dagger_i a_i a_j\ket{\psi}=0$. Then one can determine the explicit form of $\rho (\sigma_y\otimes \sigma_y)\rho^*(\sigma_y\otimes \sigma_y)$ and show that the concurrence Eq. (\ref{concurrence})  is given by
\( \mathcal{C}(\rho_{ij})=2\, \max\Big\{0,   |\langle  a^\dagger_i a^\dagger_j  \rangle   | - \langle a^\dagger_i a_i \rangle +\langle a^\dagger_ia_ia_j^\dagger a_j\rangle   \Big\}  \)
Invoking identities
\begin{eqnarray}
\bra{\psi} a^\dagger_{\vb{q}} a^\dagger_{\vb{k}}\ket{\psi} &=& -\delta_{\vb{q},-\vb{k}} \sinh\frac{\phi_{\vb{k}}}{2}\cosh\frac{\phi_{\vb{k}}}{2}; \\
\bra{\psi} a^\dagger_{\vb{q}} a_{\vb{k}}\ket{\psi}&=&\delta_{\vb{q},\vb{k}} \sinh^2\frac{\phi_{\vb{k}}}{2},
\end{eqnarray}
we obtain the explicit expression of the concurrence 
\(\mathcal{C}(\rho_{ij})=\max\{0, \mathcal{T}\}, \)
where 
\begin{multline}
\mathcal{T}=  \frac{1}{(2\pi)^d}  \bigg|\int_{\text{B.Z.}} \sinh\frac{\phi_{\vb{q}}}{2}\cosh\frac{\phi_{\vb{q}}}{2} e^{i\vb{q}\cdot (\vb{R_j} -\vb{R_i})}\, d^d \vb{q} \bigg|\\
-   \frac{1}{(2\pi)^d}  \int_{\text{B.Z.}} \sinh^2\frac{\phi_{\vb{q}}}{2}\, d^d \vb{q}. \label{complex}
\end{multline}
Here  $\text{B.Z.}$ represents the Brillouin zone.  Considering $\tanh\phi_{\vb{k}}=K/(2J \vb{k}^2+B)$ and introducing  parameters $\gamma=|\vb{R}|=|\vb{R}_i-\vb{R}_j|$, $\lambda=\sqrt{J/B}$, $\eta=|K|/B$ and $\hat{\vb{R}}=(\vb{R}_i-\vb{R}_j)/|\vb{R}|$, we can rewrite Eq. (\ref{complex}) and obtain Eq. (\ref{thermo}):

\begin{widetext}
\( \mathcal{T}(\gamma, \lambda, \eta) \approx    \frac{1}{(2\pi)^d}   \bigg|        \int_{\text{B.Z.}} d^d\vb{q}  \frac{\eta \cos(\gamma \hat{\vb{R}}\cdot \vb{q} )  }{\sqrt{  (1+ 2\lambda^2\vb{q}^2  )^2   -\eta^2     }}     \bigg|    +   \frac{1}{(2\pi)^d}   \int_{\text{B.Z.}}  d^d\vb{q} \bigg[ 1  -\frac{ 1+ 2\lambda^2\vb{q}^2    }{\sqrt{  (1+ 2\lambda^2\vb{q}^2  )^2   -\eta^2     } }  \bigg]. \)
\end{widetext}
Applying Wick's theorem to $\langle a^\dagger_ia_ia_j^\dagger a_j\rangle$ and using $|\langle a^\dagger_ia^\dagger_j\rangle|\sim \order{\eta/\lambda^2}$, $|\langle a^\dagger_i a_j\rangle| <\langle a^\dagger_i a_i\rangle\sim \order{\eta^{2+d/2}/\lambda^d}$, which are all small, we see that the quartic correlator can be neglected.


\begin{thebibliography}{56}
\bibitem [{ham()}]{hamiltonian}%
  \BibitemOpen
  \href@noop {} {}\bibinfo {note} {In general, we can write a symmetric
  exchange coupling $H_1$ as $\sum_{\langle i,j
  \rangle}(K_1S^x_iS^x_j+K_2S^y_iS^y_j+K_3S^z_iS^z_j)$, in the principal-axes
  coordinate frame. So long as $K_1\neq K_2$, this will only redefine constants
  in our discussion but not alter the analysis.}\BibitemShut {Stop}%
\bibitem [{har()}]{hardcore}%
  \BibitemOpen
  \href@noop {} {}\bibinfo {note} {The transformation preserves the Pauli
  commutation algebra only when we restrict the boson operators $\hat{a}_i$ to
  the subspace with $\hat{a}_i^\dagger \hat{a}_i\to\{0,1\}$, i.e.,
  $\hat{a}_i\rightarrow a_i\equiv P_i\hat{a}_iP_i$, with $P_i\equiv |0\rangle_i
  {}_i\langle 0|+|1\rangle_i {}_i\langle1|$. The appropriate spin operators,
  $S_i^+=a_i$ and $S_i^z=1/2-a_i^\dagger a_i$, are then, in fact, expressed in
  terms of these projected operators. Upon substituting these into the
  Hamiltonian, however, we can drop the projection operators, at the expense of
  introducing the hard-core repulsion. In the dilute-magnon limit, furthermore,
  the latter can be simply disregarded.}\BibitemShut {Stop}%
\bibitem [{unc()}]{uncertainty}%
  \BibitemOpen
  \href@noop {} {}\bibinfo {note} {We use the standard notation: $\Delta
  S^x=\sqrt{\langle (S^x- \langle S^x\rangle)^2 \rangle}$ and the minimum
  uncertainty relation is given by $\Delta S^x \Delta S^y=|\langle
  [S^x,S^y]\rangle|/2=S^z/2\approx 1/4$ when the average magnon density is
  small.}\BibitemShut {Stop}%
\bibitem [{non()}]{nonzero}%
  \BibitemOpen
  \href@noop {} {}\bibinfo {note} {Similar argument can be made for nonzero
  wavenumber modes. The parameter in the two-mode squeezing operator will
  become complex-valued.}\BibitemShut {Stop}%
\end{thebibliography}
\end{document}